\documentclass[aps,prd,preprintnumbers,showpacs]{revtex4}
\usepackage[dvips]{graphicx}
\begin{document}

%
%

\eprint{Nisho-1-2019}
\title{Nonvanishing pion masses for vanishing bare quark masses}
\author{Aiichi Iwazaki}
\affiliation{Nishogakusha University,\\ 
6-16 Sanbancho Chiyoda-ku Tokyo 102-8336, Japan.}   
\date{Jan. 10, 2019; revised Apr. 18, 2019}
\begin{abstract}
It is generally thought that pion masses vanish when bare light quark masses $m_q$ vanish. 
The pions are Nambu-Goldstone bosons of chiral $SU_{A}(2)$ symmetry.
We discuss a possibility that even when $m_q=0$,
the chiral symmetry is explicitly broken in strong coupled QCD, 
while it is not in weakly coupled QCD.
The point is that QCD monopoles are dynamical degrees of freedom in the strong coupled QCD.
We show that the monopole quark interactions break the chiral $U_5(1)$ symmetry as well as
the chiral $SU_{A}(2)$ symmetry.
They are weak relative to the other chiral symmetric 
interactions of quarks and gluons. 
These weak interactions produce small pion masses as well as quark masses $\sim 20$MeV
in the phase with the monopole condensation.
The presence of the interactions modifies the Gell-Mann Oakes Renner relation.
Using the relation we predict the value $(-\langle\bar{q}q\rangle)^{1/3}\simeq 160$MeV. 
We also show that the monopole quark interactions do not prevent
the monopole condensation in dual superconducting model. 
\end{abstract}
\hspace*{0.3cm}
\pacs{12.38.-t,12.38.Aw,11.30.Rd,14.80.Hv}

\hspace*{1cm}

\maketitle



\section{Introduction}
Pions are Nambu-Goldstone bosons associated with chiral $SU_A(2)$ symmetry
and their nonzero masses are produced by bare light quark masses.
Namely, the chiral symmetry is spontaneously broken and simultaneously chiral condensate $\langle\bar{q}q\rangle$ arises.
The condensate is the order parameter of the chiral symmetry.
It is implicitly assumed in the arguments 
that we can define the chiral symmetric strong coupled QCD when the bare light quark masses $m_q$ vanish.
Because chiral symmetric weakly coupled QCD can be defined perturbatively, it is expected that 
the chiral symmetry holds even in the strong coupled QCD. 

In this paper we discuss that the chiral symmetry does not hold in the strong coupled QCD where
QCD monopoles are supposed to be quasi-particles. We show that
the monopole quark interactions explicitly break the chiral symmetry.
We cannot define chiral symmetric strong coupled QCD even
if the bare quark masses vanish. On the other hand, weakly coupled QCD is defined to be chiral symmetric.  
We discuss the phenomenological effects of the chiral non symmetric monopole quark interactions.

\vspace{0.1cm}
It has recently been pointed out that QCD monopoles are present in strong coupled QCD and
play important roles\cite{heavy} for strong coupled QGP as well as the quark confinement\cite{dual}.
They have been discussed to be dominant components of the QGP near the transition temperature $\sim 160$MeV, while
they are absent in perturbative QCD.
Such monopoles have been also discussed\cite{iwazaki} to play a role for chiral symmetry breaking.
That is, it has been shown that the chiral non symmetric pair production of massless quarks takes place in the monopole condensed vacuum
when a classical color charge is put in the vacuum.
Furthermore, their roles for the chiral symmetry breaking have been pointed out\cite{hasegawa} in lattice gauge theory.
In particular, the association\cite{banks} with zero modes of Dirac operators has been examined:
Addition of a pair of monopole and anti-monopole to gauge field configuration in Dirac operators
increases the number of the zero modes. 
It suggests that the monopoles play a role for the generation of the chiral condensate.

\vspace{0.1cm}
In this paper we assume that the monopoles are present in strong coupled QCD and show that
the chiralities of the quarks are not conserved in the monopole quark scatterings.
It has been discussed\cite{kazama} that in order to properly define the scatterings,
we need to impose a chiral non symmetric boundary condition for quarks at the location of the monopoles.
We may describe such interactions as $(\bar{u}u+\bar{d}d)\Phi^{\dagger}\Phi$ where $u$ and $d$ denote up and down quarks, and
$\Phi$ denotes the monopole. The monopole is flavor singlet so that the quarks do not change their flavors in the scattering.
The interactions explicitly break flavor chiral $SU_{A}(2)$ and chiral $U_{A}(1)$ symmetries just similar to quark mass terms.
Obviously, the interactions generate quark masses when the monopoles condense
$\langle\Phi\rangle\neq 0$ even in the absence of bare quark masses. 

We discuss in detail the effects of the interactions in the paper. In particular, we show that the
interactions are approximately ten times smaller than the other quark-gluon interactions in QCD.
They generate the quark masses $\sim 20$MeV in the phase with the monopole condensation even when $m_q=0$.
Furthermore, we show that the chiral condensate is given by $(-\langle\bar{q}q\rangle)^{1/3}\simeq 160$MeV.
We confirm that the monopole quark interactions do not prevent the monopole condensation in dual superconducting model
for quark confinement. The fact leads to the coincidence of deconfinement transition temperature and chiral restoration transition temperature. 
Obviously, the pions are not massless even when $m_q=0$.

\section{Monopole Quark Interaction}
First, we would like to show that the chiralities are not conserved in the monopole quark scatterings. 
Namely, the monopole quark interactions break chiral $U_5(1)$ symmetry. 
In order to do so, we briefly explain monopoles in strong coupled QCD. 
We notice that the monopoles arise only in the low energy
region of QCD. In the region the coupling strength is so large
as for quarks and gluons themselves to be not appropriate dynamical variables.
It has been discussed that
the assumption of Abelian dominance\cite{abelian,maximal} may hold in the low energy region.
According to the assumption, the dynamical degrees of freedom relevant to the 
low energy physics are massless color triplet quarks, maximal Abelian ( diagonal ) gauge fields and
monopoles. Off diagonal components of the gauge fields are massive so that they are not
relevant. Although the monopoles are massive, they are nearly massless at a critical temperature where they begin to condense in vacuum.
After their condensation, the masses of excited monopoles are much less than those of off diagonal gluons. 
Thus, relevant excitations to the low energy QCD are massless quarks, Abelian ( diagonal ) gluons and the monopoles.
In the present paper we assume the Abelian dominance in order to treat explicitly the monopole excitations \cite{maedan,suga,che} in strong coupled QCD.

\vspace{0.1cm}
In $SU(3)$ gauge theory, we have three types of monopoles characterized by root vectors of $SU(3)$,
 $\vec{\epsilon}_1=(1,0), \vec{\epsilon}_2=(-1/2,-\sqrt{3}/2)$ and $\vec{\epsilon}_3=(-1/2,\sqrt{3}/2)$.
They describe the couplings with the maximal Abelian gauge fields, $A_{\mu}^{3,8}$ such as $\epsilon_i^a A_{\mu}^a$.
For example a monopole with $\vec{\epsilon}_1$ couples with $A_{\mu}^3=\epsilon_1^a A_{\mu}^a$. 
It implies that the monopole is represented by a Dirac monopole using the gauge field $A_{\mu}^3$. 
Then, the quarks coupled with the monopole
are a doublet $q=(q^+,q^-,0)$ of the color triplet. Here the index $\pm$ of $q^{\pm}$ denotes a positive or negative charged component associated
with the gauge field $A_{\mu}^3\lambda^3/2$; $\lambda^a$ are Gell-Man matrices.
Similarly the other monopoles couple with the quark doublets, $q=(q^+,0,q^-)$ and $q=(0,q^+,q^-)$. 
Thus, we consider the scattering of a monopole and a massless quark doublet $(\begin{array}{l}q^+ \\ q^-\end{array} )$. 
( The arguments below cannot be applied to the quarks in adjoint representation. ) 

\begin{equation}
\label{1}
\gamma_{\mu}(i\partial^{\mu}\mp \frac{g}{2}A^{\mu})q^{\pm}=0,
\end{equation}
where the gauge potentials $A^{\mu}$ denotes a Dirac monopole
given by 

\begin{equation}
\label{2}
A_{\phi}=g_m(1-\cos(\theta)), \quad A_0=A_r=A_{\theta}=0
\end{equation} 
where $\vec{A}\cdot d\vec{x}=A_rdr+A_{\theta}d\theta+A_{\phi}d\phi$ with polar coordinates $r,\theta$ and $\phi=\arctan(y/x)$.
$g_m$ denotes a magnetic charge with which magnetic field is given by $\vec{B}=g_m\vec{r}/r^3$.
The magnetic charge satisfies the Dirac quantization condition $g_mg=n/2$ with integer $n$ where $g$ denotes the gauge coupling of $SU(3)$ gauge theory.
Hereafter, we assume the monopoles with the magnetic charge $g_m=1/2g$.

\vspace{0.1cm}

The monopole quark ( in general, fermion ) dynamics has been extensively explored, in particular, in the situation of
monopole catalysis\cite{rubakov,callan,ezawa} of baryon decay ( so called Rubakov effect ).
We consider the scatterings of the massless quarks and the massive monopoles; 
they are sufficiently massive not to move in the low energy scatterings.
It is specific that conserved angular momentum has an additional component. That is,
it is given by

\begin{equation}
\label{3}
\vec{J}=\vec{L}+\vec{S}\mp gg_m\vec{r}/r
\end{equation}
where $\vec{L}$ ( $\vec{S}$ ) denotes orbital ( spin ) angular momentum of quark.
The additional terms $\pm gg_m\vec{r}/r$ play an important role of chiral symmetry breaking.
Owing to the term we can show that either the charge or the chirality is not conserved in the monopole quark scattering.
When the chirality ( or helicity $\sim \vec{p}\cdot\vec{S}/|\vec{p}||\vec{S}|$ ) is conserved, the spin must flip $\vec{S}\to -\vec{S}$ after the scattering 
because the momentum flips after the scattering; $\vec{p}\to -\vec{p}$.
Then, the charge must flip $g\to -g$ because of the conservation of $\vec{J}\cdot\vec{r}$, i.e.
 $\Delta(\vec{J}\cdot\vec{r})=\Delta(\vec{S}\cdot\vec{r})+\Delta(gg_mr)=0$. ( $\Delta(Q)$ denotes the change of the value $Q$ after the scattering. ) 
On the other hand, when the charge is conserved ( it leads to $0=\Delta(\vec{J}\cdot\vec{r})=\Delta(\vec{S}\cdot\vec{r}$) ),
the chirality $\vec{p}\cdot\vec{S}/|\vec{p}||\vec{S}|$ must flip because the spin does not flip $\vec{S}\to \vec{S}$.
Thus we find that either the charge or the chirality conservation is lost in the scatterings.

\vspace{0.1cm}
We cannot uniquely define the monopole quark scatterings in the Dirac equation.
To explicitly define the scatterings\cite{kazama}, 
we need to imposed a boundary condition for the quarks at the location of the monopoles.
It is either of charged conserved but chirality non conserved boundary condition or chirality conserved but charge non conserved one.
The charge is strictly conserved because of the gauge symmetry.
When the quark flips its charge, monopoles are charged or heavy charged off diagonal gluons must be produced to preserve the charge conservation.
But the processes cannot arise in the low energy scatterings. Charged monopoles are dyons and they are heavy.   	
Thus, inevitably the chirality is not conserved. The right handed quark $q_R$ is transformed to the left handed quark $q_L$ in the scatterings.
That is, 
we need to impose the boundary conditions $q^{\pm}_R(r=0)=q^{\pm}_L(r=0)$ at the location $r=0$ of the monopole.
The boundary conditions explicitly breaks the chiral $U_{A}(1)$ symmetry. 
The detailed analyses have shown\cite{rubakov,ezawa} that the chiral $U_{A}(1)$ symmetry breaking is caused by chiral anomaly in QCD.
( Even if we impose the chirality conserved but charge non conserved boundary condition $q^+_{R,L}(r=0)=q^-_{R,L}(r=0)$ at the location of the monopole, 
we can show\cite{rubakov,ezawa} that the charge is conserved,
but chirality is not conserved. 
The chirality non conservation arises from chiral condensate $\langle\bar{q}q\rangle\propto 1/r^3$
locally present around each monopole at $r=0$.
The condensate is formed by the chiral anomaly when we take into account 
quantum effects of gauge fields $A_{\mu}=\delta A_{\mu}^{quantum}+A_{\mu}^{monopole}$.  
Eventually, the chirality non conserved boundary condition is realized in physical processes. 
These results are by-products in the analyses of the Rubakov effect\cite{rubakov, ezawa}. )

Furthermore,
quarks may change their flavors in the scatterings. For instance, u quark is transformed into d quark.
Then, the monopole must have a $SU(2)$ flavor after the scatterings. But it is impossible because there are no such monopoles
with $SU(2)$ flavors in QCD; they are flavor singlet. Therefore, quarks cannot change their flavors in the scatterings with the monopole.
Quarks change only their chiralities as dictated by the boundary conditions $q^{\pm}_R(r=0)=q^{\pm}_L(r=0)$ with $q=u,d$.
It results in the symmetry breaking of the flavor $SU_A(2)$ as well as the chiral $U_A(1)$.

\vspace{0.1cm}

Quarks change their chirality without the change of their flavors in the monopole quark scatterings;
the chiralities change at the location of the monopole.
Effectively, we can describe such interactions by using monopole field $\Phi$ that
$-g'|\Phi|^2(\bar{u}_Lu_R+\bar{u}_Ru_L+\bar{d}_Ld_R+\bar{d}_Rd_L)=-g'|\Phi|^2(\bar{u}u+\bar{d}d)$ with $g'>0$.
They are not invariant under the chiral flavor $SU_A(2)$ and the chiral $U_5(1)$ transformations.
The parameter $g'$ is roughly given by the inverse of the monopole mass $M$ times 
the ratio of the coupling strength $gg_m$ to $g^2$, i.e. $g'\sim M^{-1}gg_m/g^2$. 
The monopole mass $M$ would be of the order of $\Lambda_{QCD}$.   
( More precisely, the monopole quark interactions are described by using three types of the monopole fields $\Phi_i$ ($i=1\sim 3$ ) such that 
$-g'\big(|\Phi_1|^2(\bar{q}_1q_1+\bar{q}_2q_2)+|\Phi_2|^2(\bar{q}_1q_1+\bar{q}_3q_3)+|\Phi_3|^2(\bar{q}_2q_2+\bar{q}_3q_3)\big)$
where indices $i$ of  $q_i$ denotes color component of quark $q=u$ or $d$. The interactions preserve Weyl symmetry in QCD. )

\vspace{0.1cm}

We would like to estimate the strength $g'\propto M^{-1}$ of the interactions.
We should note that the monopole quark interactions are weak compared with hadronic interactions.
The interactions do not depend on the strong coupling $\alpha_s=g^2/4\pi$ of QCD in the tree level.
This is because the quarks have color charges $g$ and the monopoles have magnetic charges $g_m=1/2g$.
Therefore, the monopole quark interactions ( $\propto g\times g_m$ ) 
does not involve $\alpha_s$ in the tree level. On the other hands, the quark gluon interactions depend on
$\alpha_s$ even in the tree level.  Numerically, the monopole quark interactions are proportional to $gg_m=1/2$, while
quark gluon or quark quark interactions are proportional to $g^2=4\pi\alpha_s$. Because $\alpha_s(Q\sim1\mbox{GeV})\sim 0.5$,
the monopole quark interactions $gg_m=1/2$ are much smaller than the other hadronic interactions $g^2\sim 2\pi$.
That is, the chiral non symmetric interactions are $10$ ( $\simeq 2\times 2\pi$ ) times smaller than the chiral symmetric interactions.
We find that $g'\sim (gg_m/g^2)M^{-1}\simeq M^{-1}/4\pi$.

\vspace{0.1cm}
We have shown that
the chiral flavor $SU_A(2)$ symmetry is explicitly broken by the monopole quark interactions even if the bare light quark masses $m_q$ vanish.
The effects of the symmetry breaking are much bigger than those of the bare mass terms.
Indeed, when the monopole condenses, the term $g'\bar{q}q|\Phi|^2$ generates 
a mass $g'\langle\Phi\rangle^2$ of quarks. We suppose that both scales of the monopole mass $M$
and the condensate $\langle\Phi\rangle$ are of the order of $\Lambda_{QCD}\sim 250$MeV. 
Then, the mass $g'\langle\Phi \rangle^2\sim 20$MeV  ( $g'\sim (gg_m/g^2)M^{-1}\simeq M^{-1}/4\pi$ ) is bigger than those of the bare quark masses ( $m_q\sim 5$MeV ). 
Therefore, the phenomena associated with the chiral symmetry breaking are mainly caused by the 
monopole quark interactions. For example, the masses of the pions are mainly determined by the interactions.

Although the above estimation is very rough, 
it seems natural that pion masses ( $\sim 140$MeV ) are 
mainly generated by the monopole quark interactions. According to the standard idea, the masses of the pions
are determined by the small bare quark masses $\sim 5$MeV. But,
it is difficult to explain the discrepancy between the pion mass $140$MeV and the quark mass $5$MeV.
On the other hand, according to our analysis, 
the pions are not Nambu-Goldstone bosons even if the bare light quark masses are absent.
Their masses are mainly given by the quark masses $20$MeV generated in the phase with the monopole condensation, 
not the bare quark masses.

\section{Monopole Condensation}
We have introduced the monopole quark interactions $-g'\bar{q}q|\Phi|^2$. 
They produce the quark masses $g'v^2$ much bigger than the bare quark masses
when the monopole condensation $\langle\Phi\rangle=v$ takes place.
Thus, the monopole condensation leads to both the quark confinement and the generation of the quark masses.
Then, we need to examine whether or not the monopole quark interactions destroy the monopole condensation.
That is, we wonder if the interactions change the potential $V(\Phi)$ of the monopole field such that 
the monopole condensation does not arise; the minimum of the potential
is given by $\Phi=0$.

In order to examine it, we evaluate the vacuum energy of the quarks coupled with the monopoles.
It is easy to evaluate the vacuum energy of the quarks $q$,

\begin{equation}
V_q(\Phi=v)=-\sum_{\vec{p},i}\sqrt{\vec{p}^2+(v^2 g')^2}
=-\frac{N\Lambda^4}{8\pi^2x^4}\Big(x(x^2+1)^{3/2}-\frac{1}{2}x(x^2+1)^{1/2}-\frac{1}{2}\log(x+(x^2+1)^{1/2})\Big)
\end{equation}
with $x\equiv|\Lambda/v^2 g'|\ge 0$,
where $N$ denotes quark's internal degrees of freedom,
that is, $N=2\times 2\times 3=12$ ( spin$\times$ flavor $\times$ color ).
We take a cut off parameter $\Lambda$ in the integral over the range of $|\vec{p}|$. The parameter should be 
in a range $(500\sim 1000)$MeV. Beyond the value, the assumption of Abelian dominance does not hold, or
the monopole excitations are absent. 

%
%
%
%

We assume the potential $V_0(\Phi)$ of the monopole in dual superconducting model;
$V_0(\Phi)=-\mu^2|\Phi|^2+\lambda|\Phi|^4$ with $\mu^2>0$ and $\lambda>0$.
The model describes quark confinement, which is realized by the monopole condensation $\langle\Phi\rangle\neq 0$.
The minimum of the potential is given by $\langle\Phi\rangle=\sqrt{\mu^2/2\lambda}$.

Now, we examine the minimum of the potential $V(\Phi)=V_0(\Phi)+V_q(\Phi)$.
We rewrite the potential $V_0(\Phi)$ in terms of the variable $x$,

\begin{equation}
V(\Phi)=\Lambda^4(-\frac{z}{x}+\frac{w}{x^2})
\end{equation}
with $z\equiv \mu^2/(g'\Lambda^3)>0$ and $w\equiv \lambda/(g'^2\Lambda^2)>0$.

Without the monopole quark interaction, the monopoles condense, $\langle\Phi\rangle=\sqrt{\mu^2/2\lambda}$.
But it is not obvious for the monopoles to condense when the monopole quark interactions are switched on.
We can show that the potential $V_q(\Phi)+V_0(\Phi)$ with $N=12$ has the minimum at nonzero $x>0$ ( $\Phi\neq 0$ ).
This is because
$V_q+V\to \Lambda^4w/x^2$ as $x\to 0$ ( $\Phi\to \infty$ ) 
and $V_q+V\to -3\Lambda^4/(2\pi^2) -\Lambda^4z/x$ as $x\to +\infty$ ( $\Phi\to 0$ ). 
Therefore, we find the presence of the minimum at nonzero $x=|\Lambda/v^2g'|$, that is, nonzero $\Phi=v$.
Even if the monopole quark interactions are switched on, the confinement mechanism by the monopole condensation
still remain; the monopole condensation takes place. 
 
Consequently, the monopole quark interactions introduced in the present paper do not destroy the monopole condensation so that they
generate the masses of the quarks in the monopole condensed phase.
In the phase the chiral symmetry is explicitly broken. 
 
The result is consistent with our previous result\cite{iwazaki}. 
Namely, the chiral non symmetric pair production $\langle dQ_5/dt\rangle \neq 0 $ takes place in the 
monopole condensed vacuum $\langle \Phi\rangle \neq 0$ when an external classical color charge is put in the vacuum.
$Q_5$ denotes the difference of the number of the quarks with positive ( negative ) chirality; $Q_5=N_{+}-N_{-}$. 
Obviously the chiral symmetric pair production $\langle dQ_5/dt\rangle =0 $
takes place in a vacuum without the monopoles when the classical charge is put in the vacuum.
It is the standard Schwinger mechanism. 
Therefore, we find that the generation of the quark masses caused by the monopole condensation is
consistent with our previous result of the chiral asymmetric pair production in the vacuum with the condensation.

We mention that our result $\langle dQ_5/dt\rangle \neq 0 $ is derived 
by using the anomaly equations; $\partial_{\mu}J_5^{\mu}\propto \vec{E}\cdot\vec{B}$.
The anomaly also leads to the presence of the monopole quark interactions, that is, the boundary conditons for the quarks at the location of the monopole. 
( Such a chirality production\cite{iwazaki2,fukushima} has been phenomenologically discussed 
in the decay of glasma produced just after high energy heavy ion collisions. )

\section{Gell-Mann Oakes Renner relation}
Here we would like to mention that the Gell-Mann Oakes Renner relation is modified in the presence of the monopole quark interactions
in the following,

\begin{equation}
\label{6}
m_{\pi}^2f_{\pi}^2=-2(m+m_q)\langle\bar{q}q\rangle \quad  \mbox{with} \quad \langle\bar{q}q\rangle=\langle\bar{u}u\rangle=\langle\bar{d}d\rangle
\end{equation}
where $m\equiv v^2g'$ denotes the quark mass generated by the monopole condensation.
We should note that $m\simeq 20$MeV and $m_q\simeq 5$MeV.

 On the other hand, the vacuum expectation value $\langle\bar{q}q\rangle$ is estimated by using the free field of the quark with mass $m+m_q$,
 
 \begin{equation}
 \label{7}
 \langle\bar{q}q\rangle=-\frac{3\Lambda^3}{2\pi^2x^3}\Big(x\sqrt{x^2+1}-\log(x+\sqrt{x^2+1})\Big),
 \end{equation}
with $x=\Lambda/(m+m_q)$.
Using the observed values $m_{\pi}=140$MeV and $f_{\pi}=93$MeV, we numerically solve both equations (\ref{6}) and (\ref{7}) for $x$ with the use of
the quark mass $m=20$MeV. Then we find $x\simeq 38$ or the cutoff parameter $\Lambda=950$MeV.
The value of the cutoff parameter is reasonable because 
the parameter is such one that it distinguishes the energy region of the strong coupled QCD from that of the weakly coupled QCD.
The strong coupled QCD is realized in the energy region below the value. 

It also turns out\cite{miyamura} that the monopole condensation gives rise to the chiral condensation $\langle\bar{q}q\rangle$ because of the generation of the quark masses.
We predict the value $(-\langle\bar{q}q\rangle)^{1/3}\simeq 160$MeV, based on
the modified Gell-Mann Oakes Renner relation in eq(\ref{6}) and the value $m=20$MeV.

\section{Conclusion}
We have analyzed the monopole quark interactions assuming that the monopoles are relevant variables in the strong coupled QCD. 
We have found that the chiralities of the quarks are not conserved in the monopole quark scatterings.
The chiral non conservation takes place
at the location of the monopoles. That is, the quarks change their chiralities at the location of the monopoles.
We effectively describe such a interaction as $g'\bar{q}q\Phi^{\dagger}\Phi$ whose strength
has been shown to be relatively weak compared with the other quark gluon interactions; $g'\simeq \Lambda_{QCD}^{-1}/4\pi$.
The interactions generate quark masses $m\simeq 20$MeV when the monopoles condense, $\langle\Phi\rangle\neq 0$.
Therefore, the chiral symmetry is explicitly broken in the strong coupled QCD even if the bare light quark masses $m_q$ vanish. 
Pions are not Nambu-Goldstone bosons even when $m_q=0$. 
The small pion masses are mainly caused by the monopole quark interactions, not small bare quark masses $m_q$.
The result could be examined in lattice gauge theory.

We have also shown that the monopole quark interactions do not destroy the monopole condensation.
The minimum of the monopole potential
is non trivial even if the interactions are taken into account.

The monopole quark interactions modify the Gell-Mann Oakes Renner relation such that
$m_{\pi}^2f_{\pi}^2=-2(m+m_q)\langle\bar{q}q\rangle$. It
shows that the pions are not massless
even if the bare quark masses vanish. We have numerically confirmed that the modified Gell-Mann Oakes Renner relation
is satisfied when the cutoff parameter $\Lambda$ used in the evaluation of $\langle\bar{q}q\rangle$ is taken such as $\Lambda\simeq 1000$MeV.
The parameter distinguishes the energy region of the strong coupled QCD from the one of the weakly coupled QCD.
We have also found $(-\langle\bar{q}q\rangle)^{1/3}\simeq 160$MeV from the relation.

\vspace{0.1cm}
The monopole condensation causes quark confinement by producing color electric flux tube between a quark and an antiquark.
If the chiral symmetry is not broken in the confinement phase, the production of the massless quark pairs makes the flux tube
decay and destroys the quark confinement. But the massless quarks acquire their masses $g'\langle\Phi\rangle^2$ from the monopole condensation.
Therefore, the quark pair production does not arise even when the flux tube is formed as long as its length is short.
Our result supports the coincidence of the phase transition temperatures in the quark deconfinement and the chiral symmetry restoration. 
( As we noticed, the coincidence arises only for quarks in color triplet. In the case of quarks in color octet, the coincidence
does not hold. )   
Therefore, we find that the proposed monopole quark interactions in the strong coupled QCD gives a consistent picture of the quark confinement
and chiral symmetry breaking, although the standard idea such as pions being Nambu-Goldstone bosons does not hold.

 \vspace{0.2cm}
The author
expresses thanks to Prof. O. Morimatsu, KEK theory center for useful comments
and discussions.



\end{document}